\documentclass[aps,twocolumn,prl]{revtex4}
\usepackage{graphicx}

\usepackage{bm}
\usepackage{amsmath}
\usepackage{amsfonts}

\newcommand{\tr}{{\rm Tr}}
\newcommand{\ket}[1]{| #1 \rangle}
\newcommand{\bra}[1]{\langle #1 |}
\newcommand{\braket}[2]{\langle #1 | #2 \rangle}

\begin{document}
\title{Quantum nonlocality without entanglement in a pair of qubits}

\author{Masato Koashi, Fumitaka Takenaga, Takashi Yamamoto, and Nobuyuki Imoto}
\affiliation{Division of Materials Physics, 
Department of Materials Engineering Science,
Graduate School
of Engineering Science, Osaka University, 
1-3 Machikaneyama, 
Toyonaka, Osaka 560-8531, Japan}
\affiliation{CREST Photonic Quantum Information Project, 4-1-8 Honmachi, Kawaguchi, Saitama 331-0012, Japan}

\begin{abstract} 
We consider unambiguous discrimination of two separable bipartite
 states, 
one being pure and the other being a rank-2 mixed state.
There is a gap between the optimal success probability 
under global measurements and the one achieved by generalized measurements
with separable measurement operators. We show that even the latter 
success probability cannot be achieved via
local operations and classical communication, leaving a nonzero gap in between.

\end{abstract}

\maketitle

In the recent progress on understanding of quantum information,
distinction between classical interaction and quantum interaction
has played a key role. 
For a bipartite system AB, a
 restricted class of operations called 
LOCC (local operations and classical communication) is considered,
and any state that can be prepared from scratch under LOCC is 
called separable. The rest of the states are termed entangled 
states, which are prepared only through quantum interaction between 
A and B. 
One might expect that the classical/quantum 
boundary would be clear-cut in this paradigm, and separable states 
would have nothing to do with the quantum interaction. 
But that is not the case.
Peres and Wootters 
suggested \cite{Peres-Wootters91}
that a set of bipartite separable states can be distinguished better 
by a global measurement, which can be realized only by quantum 
interaction between the two systems. 
Later, using different sets of states and different measures of 
distinguishability,
a stronger evidence was shown \cite{Massar-Popescu95}, 
a nonzero gap between the optimal 
global distinguishability $P^{(\rm glo)}_{\rm opt}$ and 
the LOCC counterpart $P^{(\rm LOCC)}_{\rm opt}$ 
was found \cite{BDFMRSSW99}, 
and more
recently, global measurements surpassing theoretically derived 
$P^{(\rm LOCC)}_{\rm opt}$ were 
experimentally demonstrated \cite{POWB05,NAACGNS06}.

If a measurement is implementable under LOCC, it must be
a {\em separable measurement}, for which every 
POVM (positive-operator-valued measure) element
is separable (written as a sum of positive product operators
like $\sum_j \hat{A}_j\otimes \hat{B}_j$). 
In most of the above examples, the measurement 
achieving $P^{(\rm glo)}_{\rm opt}$ is inseparable,
and hence one can use it  
to create entanglement from a separable state \cite{TDL01}. 
The reason why the measurement cannot be carried out 
under LOCC is obvious in this respect. An exception is the 
example given in \cite{BDFMRSSW99}, composed of a cleverly chosen set of 
nine (or eight) pure orthogonal 
 states of a pair of qutrits (three level systems).
In this case, the best result is achieved by a separable measurement.
Nonetheless, the LOCC measurements are proved to be less efficient
and there is a nonzero gap between 
$P^{(\rm sep)}_{\rm opt}$ and $P^{(\rm LOCC)}_{\rm opt}$,
where $P^{(\rm sep)}_{\rm opt}$ is the optimal distinguishability 
among separable measurements.
This implies that the optimal separable measurement 
has a striking irreversibility under LOCC: 
we must invest entanglement 
to implement the measurement, whereas we can never use it 
to create entanglement from scratch. 
We see interesting parallels between this and 
a bound entangled state \cite{HHH98}. For the latter, 
we must invest entanglement 
to create the state, whereas we can never extract 
from it a standard form of entanglement 
like a singlet state of two spin-$1/2$ particles. 
In comparison to bound entanglement, our understanding 
about such bizarre measurements or processes are 
still rudimentary, including the very reason why the LOCC
cannot implement those. For example, we know that the 
bound entanglement is unique to higher-level systems and
does not appear in a qubit-pair system \cite{HHH97}, 
but we know little 
about the same question for
$P^{(\rm sep)}_{\rm opt}>P^{(\rm LOCC)}_{\rm opt}$.

The aim of this paper is to provide 
a discrimination problem exhibiting such a phenomenon,
which is much simpler in many ways: the system is a qubit 
pair, and the number of separable states to distinguish 
is just two. The description of the two states involve
only three pure product states. The two states may look 
rather mundane and with little contraption, but we can still 
prove $P^{(\rm glo)}_{\rm opt}>
P^{(\rm sep)}_{\rm opt}>P^{(\rm LOCC)}_{\rm opt}$. 
The key element in the proof
is the observation that refinement process of 
any LOCC measurement must proceed
alternately between the two systems.
This should be contrasted to the property used
in the proof of \cite{BDFMRSSW99}, that the 
refinement process can always be regarded as a
continuous one.

The problem we consider here is the unambiguous 
discrimination \cite{Ivanovic87}
 between the two separable states of a two-qubit system defined by
\begin{eqnarray}
 \hat{\rho}_0 &=& \ket{00}\bra{00},
 \\
 \hat{\rho}_1 &=& (1/2)(\ket{++}\bra{++}+\ket{--}\bra{--}),
\end{eqnarray}
where $\ket{\pm}\equiv 2^{-1/2}(\ket{0}\pm \ket{1})$.
The two qubits are secretly 
prepared in one of the two states before 
one qubit is sent to Alice and the other one
to Bob. They should determine the identity of 
the prepared state by a quantum 
measurement. We allow them to declare that the 
measurement has been a failure, but otherwise they 
must identify the state with no errors. 
Such a measurement is generally described by 
a POVM 
$\{\hat{F}_m\}_{m=0,1,2}$ satisfying
\begin{eqnarray}
 \tr(\hat{F}_0\hat{\rho}_1)=\tr(\hat{F}_1\hat{\rho}_0)=0,
\end{eqnarray}
where the outcome $m=2$ corresponds to the failure and 
$m=0,1$ to the identification of state $\hat{\rho}_m$.
For $j=0,1$, let $\gamma_j\equiv \tr(\hat{F}_j\hat{\rho}_j)$ 
be the probability of success when the prepared state was 
$\hat{\rho}_j$. What we seek is the supremum of $\gamma_1$
over all the allowed measurements with a fixed value of 
$\gamma_0$, which we will denote by $P_{\rm opt}(\gamma_0)$.
A related problem is the optimization of the averaged 
success rate $\gamma_{\rm ave}\equiv \eta_0 \gamma_0 + \eta_1 \gamma_1$
when $\rho_j$ is prepared with probability $\eta_j$.
The optimized value $Q_{\rm opt}$ can be straightforwardly calculated once
we learn the function $P_{\rm opt}(\gamma_0)$, namely,
$Q_{\rm opt}
=\max_{\gamma} [\eta_0 \gamma + \eta_1 P_{\rm opt}(\gamma)]$.

\begin{figure}
\center{\includegraphics[width=.95\linewidth]{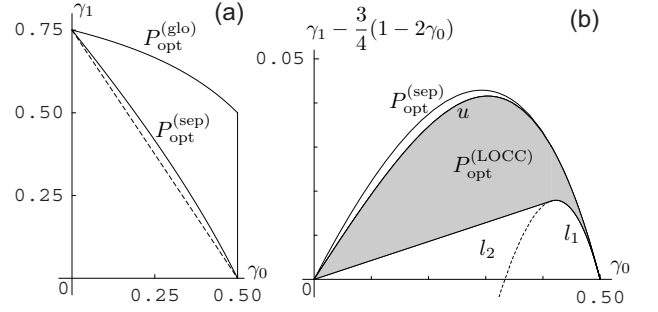}}
\caption{(a) Bounds on success probabilities $(\gamma_0,\gamma_1)$.
(b) Expanded after taking the dotted straight line in (a) as the
base line. $P^{({\rm LOCC})}_{\rm opt}$ should lie in the shaded region.
\label{fig:1}
}
\end{figure}

If we allow Alice and Bob arbitrary global quantum operations, 
the optimal probability 
$\gamma_1=P_{\rm opt}^{(\rm glo)}(\gamma_0)$ is known 
\cite{BHH03} to 
be achieved by the following protocol. First, they globally 
measure whether their bit values coincide, namely,
project the state to the subspace spanned by 
$\{\ket{00},\ket{11}\}$ and to the one by 
$\{\ket{01},\ket{10}\}$. If 
the prepared state was $\hat{\rho}_1$,
the latter case occurs with probability 1/2
and then the measurement successfully identifies it. 
Otherwise, they are left with the discrimination between 
$\ket{\psi_0}\equiv \ket{00}$
and $\ket{\psi_1}\equiv 
2^{-1/2}(\ket{00}+\ket{11})$
with $s\equiv |\braket{\psi_0}{\psi_1}|^2=1/2$.
 This is a well-known problem 
and $\ket{\psi_0}$ can be identified with probability 
$\gamma_0(\le 1/2)$ while $\ket{\psi_1}$ can be identified with
$1-s/(1-\gamma_0)$ \cite{Jaeger-Shimony95}. The whole protocol achieves 
$\gamma_1$ equal to 
\begin{eqnarray}
 P_{\rm opt}^{\rm (glo)}(\gamma_0)=
1-[4(1-\gamma_0)]^{-1}
\end{eqnarray}
for $0\le \gamma_0 \le 1/2$, which
is plotted in Fig.~\ref{fig:1} (a).

Next, we determine 
the optimal probability  
$\gamma_1=P_{\rm opt}^{({\rm sep})}(\gamma_0)$ 
among separable measurements.
It is convenient to use the high symmetry of states 
$\hat\rho_0$ and $\hat{\rho}_1$. These states are invariant 
under any of the following maps --- Exchange of the two qubits,
given by the map $S^{(1)}: 
\ket{ij}\bra{kl}\mapsto \ket{ji}\bra{lk}$;
simultaneous phase flip, given by 
 $S^{(2)}: 
\ket{ij}\bra{kl}\mapsto (-1)^{i+j+k+l}\ket{ij}\bra{kl}$;
(global) transpose, given by  
$S^{(3)}: 
\ket{ij}\bra{kl}\mapsto \ket{kl}\bra{ij}$;
and partial transpose, given by 
$S^{(4)}: 
\ket{ij}\bra{kl}\mapsto \ket{il}\bra{kj}$.
As a result, if a separable measurement $\{\hat{F}_m\}$ achieves 
success probabilities $(\gamma_0, \gamma_1)$, 
all the POVMs generated by applying the above maps
are physically valid (note the separability of $\hat{F}_m$
for $S^{(4)}$ \cite{Peres96})
and give the same probabilities $(\gamma_0, \gamma_1)$.
Then, the symmetrized measurement, which executes one of those $2^4$
measurements randomly, also achieves the same $(\gamma_0, \gamma_1)$.
Let us define the symmetrizing map via the composition
${\cal S}\equiv \prod_{n=1}^4 2^{-4}(I+S^{(n)})$, where $I$ is the 
identity map. The POVM of the symmetrized measurement is given by 
$\{{\cal S}(\hat{F}_m)\}$. Any symmetrized self-adjoint 
operator has a 
simple form with four real parameters 
in the matrix representation on the basis 
$\{\ket{00}, \ket{01},\ket{10},\ket{11}\}$ as 
\begin{eqnarray}
[a,b,c; \mu]\equiv
\left(
 \begin{array}{cccc}
  a & 0 & 0 & \mu\\
  0 & c & \mu & 0\\
  0 & \mu & c & 0\\
  \mu & 0 & 0 & b\\
\end{array}
\right).
\end{eqnarray}
It is positive (semidefinite) if and only if 
$a\ge 0$, $ab\ge \mu^2$, and $c\ge |\mu|$.
With this notation, the two initial states are
written by $\hat{\rho}_0=[1,0,0;0]$ and 
$\hat{\rho}_1=(1/4)[1,1,1;1]$.

The conditions $\tr[\hat{\rho_0}{\cal S}(\hat{F}_0)]=\gamma_0$
and $\tr[\hat{\rho_1}{\cal S}(\hat{F}_0)]=0$ require that 
${\cal S}(\hat{F}_0)=\gamma_0[1,1,1;-1]$.
To satisfy $\tr[\hat{\rho_0}{\cal S}(\hat{F}_1)]=0$ and 
to be positive, ${\cal S}(\hat{F}_1)$
must be written as ${\cal S}(\hat{F}_1)=[0,b,c;0]$.
The positivity of 
${\cal S}(\hat{F}_2)=[1-\gamma_0, 1-\gamma_0-b,1-\gamma_0-c;\gamma_0]$
requires $c\le 1-2\gamma_0$ and $b \le (1-2\gamma_0)/(1-\gamma_0)$.
The optimal value of $\gamma_1=\tr[\hat{\rho_1}{\cal S}(\hat{F}_1)]
=(b+2c)/4$ is thus given by
\begin{eqnarray}
 P_{\rm opt}^{\rm (sep)}(\gamma_0)=
1-\gamma_0-[4(1-\gamma_0)]^{-1}
\end{eqnarray}
for $0\le \gamma_0 \le 1/2$, which is shown
in Fig.~\ref{fig:1}.
This is achieved 
if and only if ${\cal S}(\hat{F}_m)=\hat{F}_m^{\rm (sep)}$ with
\begin{eqnarray}
 \hat{F}_0^{\rm (sep)} &\equiv& 2\gamma_0 (
\hat{P}_+\otimes\hat{P}_-+\hat{P}_-\otimes\hat{P}_+)
\\
 \hat{F}_1^{\rm (sep)} &\equiv& (1-2\gamma_0)(\hat{P}_0\otimes\hat{P}_1
+\hat{P}_0\otimes\hat{P}_1)
 \nonumber \\
  && + (1-\xi_0) \hat{P}_1\otimes\hat{P}_1
 \\
\hat{F}_2^{\rm (sep)} &\equiv& [(1+\xi_0)/2]
(\hat{P}_{\gamma +}\otimes\hat{P}_{\gamma +}
+\hat{P}_{\gamma -}\otimes\hat{P}_{\gamma -})
\label{eq:F2sep}
\end{eqnarray}
where $\xi_0\equiv \gamma_0/(1-\gamma_0)$, 
$\ket{\gamma\pm}\equiv \sqrt{1-\gamma_0}\ket{0}\pm
\sqrt{\gamma_0}\ket{1}$,
and we denoted $\hat{P}_X\equiv \ket{X}\bra{X}$.

Next, for $0<\gamma_0<1/2$, 
we will show that the optimal separable measurement 
$\{\hat{F}_m^{\rm (sep)}\}$ cannot be implemented under LOCC,
and there is a nonzero gap between $P_{\rm opt}^{\rm (sep)}$ and
$P_{\rm opt}^{\rm (LOCC)}$.
In a general LOCC measurement protocol, Alice first obtains an outcome
$i_1$ and tells Bob. If they stopped the protocol here, it would be
regarded as a measurement with POVM $\{\hat{G}_{i_1}\}_{i_1}$ with 
$\hat{G}_{i_1}=\hat{A}_{i_1}\otimes \hat{1}$. 
Next, Bob obtains an outcome $i_2$ and tells Alice. 
The POVM is refined to $\{\hat{G}_{i_1i_2}\}_{i_1,i_2}$ with 
$\hat{G}_{i_1i_2}=\hat{A}_{i_1}\otimes \hat{B}_{i_1i_2}$
at this point, where $\sum_{i_2}\hat{G}_{i_1i_2}=\hat{G}_{i_1}$.
Repeating such a process $n$ rounds,
they carry out a measurement with POVM  
$\{\hat{G}_{i_1\ldots i_n}\}_{i_1,\ldots,i_n}$, with 
$\hat{G}_{i_1\ldots i_n}=
\hat{A}_{i_1i_2\cdots}\otimes \hat{B}_{i_1i_2\cdots}$.
Finally, they determine the final outcome $m=0,1,2$ 
according to a rule $m=\Omega(i_1,\ldots,i_n)$.

\begin{figure}
\center{\includegraphics[width=.5\linewidth]{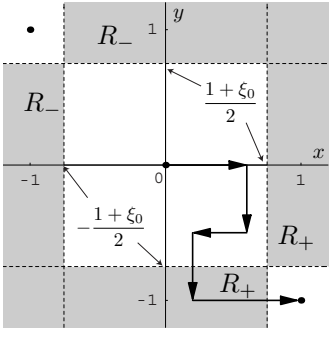}}
\caption{Refining process of an LOCC measurement.
\label{fig:xy}
}
\end{figure}

Let us introduce a ``weight'' of a POVM element $\hat{G}$
by $w(\hat{G})\equiv \bra{00}\hat{G}\ket{00}=\tr(\hat\rho_0\hat{G})$. 
As long as $w>0$, 
we can keep track of the progress 
$\hat{1}\otimes\hat{1}\to$ $\hat{G}_{i_1}\to$ $\hat{G}_{i_1i_2}\to \cdots$
$\to\hat{G}_{i_1\ldots i_n}$ using the two real parameters 
defined by 
\begin{eqnarray}
 x(\hat{A})\equiv {\rm Re}\; A_{01}/A_{00}, \;
 y(\hat{B})\equiv {\rm Re}\; B_{01}/B_{00},
\end{eqnarray}
where $A_{jk}\equiv \bra{j}\hat{A}\ket{k}$ and 
$B_{jk}\equiv \bra{j}\hat{B}\ket{k}$.
The starting point for $\hat{1}\otimes\hat{1}$ is 
$(x(\hat{1}), y(\hat{1}))=(0,0)$. This is followed by the point
$(x(\hat{A}_{i_1}), 0)$, representing $\hat{G}_{i_1}$,
and then the point $(x(\hat{A}_{i_1}), y(\hat{B}_{i_1i_2}))$ for 
$\hat{G}_{i_1i_2}$, and so on.
We notice that either $x$ or $y$, and not both, 
changes in a single step. Hence the trajectory on the $xy$ plane
is a zigzag line as in Fig.~\ref{fig:xy}.
This property, reflecting the fact that
  Alice and Bob can refine
the measurement operator {\em only alternately}
to arrive at the final outcome, is 
crucial in
showing that the optimal separable measurement is not feasible.

The parameters $w,x,y$ take distinct values for $m=0,1$.
If $m=\Omega(i_1,\ldots,i_n)=0$ and hence
$\tr(\hat{G}_{i_i\ldots i_n}\hat{\rho}_1)=0$,
then $\hat{G}_{i_1\ldots i_n}$ should be proportional to 
$\hat{P}_+\otimes\hat{P}_-$
or to $\hat{P}_-\otimes\hat{P}_+$, namely,
\begin{eqnarray}
 (x,y)=(1,-1) \;\;{\rm or} \;\; (-1,1) \;\; {\rm for}\;\; m=0.
\label{eq:condm0}
\end{eqnarray}
If $m=\Omega(i_1,\ldots,i_n)=1$, 
$\tr(\hat{G}_{i_i\ldots i_n}\hat{\rho}_0)=0$ means that 
\begin{eqnarray}
 w(\hat{G}_{i_i\ldots i_n})=0 \;\; {\rm for}\;\; m=1.
\label{eq:condm1}
\end{eqnarray}

In order to achieve $\gamma_1=P_{\rm opt}^{\rm (sep)}(\gamma_0)$,
the outcomes with $m=\Omega(i_1,\ldots,i_n)=2$ must also have 
distinct values for $(x,y)$, namely,
\begin{eqnarray}
 (x,y)=(\sqrt{\xi_0},\sqrt{\xi_0})
 \;{\rm or} \; (-\sqrt{\xi_0},-\sqrt{\xi_0}) \;\; {\rm for}\;\; m=2.
\label{eq:condm2}
\end{eqnarray}
This is because we see from Eq.~(\ref{eq:F2sep})
that the range of 
${\cal S}(\hat{F}_2)=\hat{F}^{({\rm sep})}_2$ is 
spanned by $\{\ket{\gamma+}\ket{\gamma+}, \ket{\gamma-}\ket{\gamma-}\}$
and includes no other product states since $\gamma_0>0$.
Hence $\hat{G}_{i_1\ldots i_n}$ should be proportional to 
$\hat{P}_{\gamma\pm}\otimes\hat{P}_{\gamma\pm}$.
This allows us to find a pair of linear functionals $f_{\pm}$
that are nonpositive for any final outcome $\hat{G}_{i_1\ldots i_n}$.
These are defined by 
\begin{eqnarray}
 f_\pm(\hat{F})\equiv  F_{0011}+F_{0110}+F_{1001}+F_{1100} -4\xi_0F_{0000}
\nonumber \\
 \pm (1+\xi_0)(F_{0010}+F_{1000}-F_{0001}-F_{0100}),
\end{eqnarray}
where $F_{ijkl}\equiv \bra{ij}\hat{F}\ket{kl}$. For a
product operator $\hat{G}=\hat{A}\otimes\hat{B}$
with $x(\hat{A})=x$ and $y(\hat{B})=y$, we have 
\begin{eqnarray}
 f_\pm(\hat{G})&=& w(\hat{G})[4(xy-\xi_0) \pm 2(1+\xi_0)(x-y)]
\label{eq:fpmG1}
\\
&=& w(\hat{G})
[\chi_\pm(x,y)+(1-\xi_0)^2],
\label{eq:fpmG2}
\end{eqnarray}
where 
$\chi_\pm(x,y)\equiv [2x\mp (1+\xi_0)][2y\pm (1+\xi_0)]$.
Comparing Eqs.~(\ref{eq:condm0})--(\ref{eq:condm2}) and
(\ref{eq:fpmG1}),
we conclude that $f_{\pm}(\hat{G}_{i_1\ldots i_n})$ 
should be either 0 or $-8(1+\xi_0)w$ in order to
achieve $\gamma_1=P_{\rm opt}^{\rm (sep)}(\gamma_0)$.
As we will see, this is impossible with LOCC.

Since $\gamma_0>0$, there is an outcome 
with $m=0$ satisfying $w(\hat{G}_{i_1 \cdots i_n})>0$.
From Eq.~(\ref{eq:condm0}), we may assume that 
$(x,y)=(1,-1)$ for this outcome (since the case for $(-1,1)$
is similar). Let us define regions 
$R_{\pm}\equiv \{(x,y)|\chi_\pm(x,y)\ge 0\}$, which 
are shown by shaded areas in Fig.~\ref{fig:xy}.
It is obvious that the zigzag path
from the origin
must land on $R_+$ at least once 
in order to reach $(1,-1)$.
Let $\hat{G}_{i_1 \cdots i_m}$ be the point on 
$R_+$. From Eq.~(\ref{eq:fpmG2}), we see
$f_+(\hat{G}_{i_1 \cdots i_m})\ge 
w(\hat{G}_{i_1 \cdots i_m})(1-\xi_0)^2
\ge w(\hat{G}_{i_1 \cdots i_n})(1-\xi_0)^2>0$ since 
$\gamma_0<1/2$. This implies that there exists at least
one final outcome satisfying 
$f_+(\hat{G}_{i_1 \cdots i_m j_{m+1}\cdots j_n})>0$.
Hence we conclude that no finite-round LOCC protocols achieve 
$\gamma_1=P_{\rm opt}^{\rm (sep)}(\gamma_0)$.

Since the above reasoning still allows for the 
existence of LOCC protocols achieving $\gamma_1$
arbitrarily close to $P_{\rm opt}^{\rm (sep)}(\gamma_0)$,
we need a more quantitative argument 
to show that there is a nonzero gap 
$P_{\rm opt}^{\rm (sep)}(\gamma_0)- 
P_{\rm opt}^{\rm (LOCC)}(\gamma_0)>0$. First,
we divide all the possible final outcomes $(i_1,\ldots,i_n)$ 
into three groups $\Gamma_0, \Gamma_+, \Gamma_-$ 
according to the trajectories
$\hat{1}\otimes\hat{1}\to$ $\hat{G}_{i_1}\to$ $\hat{G}_{i_1i_2}\to \cdots$
$\to\hat{G}_{i_1\ldots i_n}$ by the following rules.
(i) $(i_1,\ldots,i_n)\in\Gamma_0$
if the trajectory never goes into 
the region $R_+\cup R_-$ as long as the weight $w$ is positive.
(ii) $(i_1,\ldots,i_n)\in\Gamma_+$
if the first point of trajectory in the region 
$R_+\cup R_-$ is in the region $R_+$.
(iii) $(i_1,\ldots,i_n)\in\Gamma_-$ otherwise. 
It is obvious that in the last case, 
the first point of trajectory in the region 
$R_+\cup R_-$ is in $R_-$.

Let $\hat{K}^{(\pm)}_m$ be the sum of the final POVM elements
$\hat{G}_{i_1\ldots i_n}$ satisfying 
$(i_1,\ldots,i_n)\in\Gamma_\pm$ and 
$\Omega(i_1,\ldots,i_n)=m$. Define
$\hat{K}^{(\pm)}=\hat{K}^{(\pm)}_0+\hat{K}^{(\pm)}_1+\hat{K}^{(\pm)}_2$.
Since $\hat{K}^{(+)}$ is also written as the sum of operators 
$\hat{G}_{i_1\ldots i_m}$ for the first points entering $R_+$,
we have $f_+(\hat{K}^{(+)})\ge w(\hat{K}^{(+)})(1-\xi_0)^2$
from Eq.~(\ref{eq:fpmG2}). On the other hand,
we see $f_+(\hat{K}^{(+)}_0)\le 0$ and 
$w(\hat{K}^{(+)}_1)=f_+(\hat{K}^{(+)}_1)=0$
from Eqs.~(\ref{eq:condm0}), (\ref{eq:condm1}), and (\ref{eq:fpmG1}).
Then we have
\begin{eqnarray}
 f_\pm(\hat{K}^{(\pm)}_2)\ge [w(\hat{K}^{(\pm)}_0)
+w(\hat{K}^{(\pm)}_2)](1-\xi_0)^2,
\label{eq:fpmK2}
\end{eqnarray}
where a similar result for $\Gamma_-$ is included.
As we have seen before, no outcome for $m=0$ belongs to $\Gamma_0$.
Hence $\hat{K}^{(+)}_0+\hat{K}^{(-)}_0=\hat{F}_0$ and
\begin{eqnarray}
 w(\hat{K}^{(+)}_0)+w(\hat{K}^{(-)}_0)=\gamma_0.
\label{eq:kpmgamma}
\end{eqnarray}

The next step is to quantify how much penalty 
is imposed on $\gamma_1$ when 
Eq.~(\ref{eq:condm2}) is not satisfied.
Let $\hat{N}$ be a positive separable operator.
We will prove that if $\hat{F}_2-\hat{N}$ is positive,
$\gamma_1 \le P_{\rm opt}^{\rm (sep)}(\gamma_0)- g(\hat{N})$
where $g(\hat{N})\equiv
(1/4)\tr\{\hat{N}[\xi_0^2,1,1; -(1+\xi_0)/2]\}$.

Let ${\cal S}(\hat{N})=[a',b',c';\mu']$ and
${\cal S}(\hat{F}_1)=\hat{F}_1^{\rm (sep)}-[0,\Delta b,\Delta c; 0]$.
${\cal S}(\hat{F}_2-\hat{N})$
$=\hat{F}_2^{\rm (sep)}-{\cal S}(\hat{N})+[0,\Delta b,\Delta c; 0]$
$=[1-\gamma_0-a', \gamma_0\xi_0 -b'+\Delta b, 
\gamma_0-c'+\Delta c; \gamma_0-\mu']$
must be positive. This leads to (i) $\Delta c\ge c'-\mu'$, which is 
obvious, and (ii) $\Delta b \ge \xi_0^2 a' + b' - 2 \xi_0 \mu'$,
which is proved as follows. When $a'=1-\gamma_0$, 
$\mu'=\gamma_0$ must hold and then (ii) is true since
$\gamma_0\xi_0 -b'+\Delta b\ge 0$. When $a'<1-\gamma_0$,
assuming that (ii) is false would lead to 
\begin{eqnarray}
 (1-\gamma_0-a')(\gamma_0\xi_0 -b'+\Delta b)
-(\gamma_0-\mu')^2
\nonumber \\
<-(\xi_0a'-\mu')^2\le 0,
\end{eqnarray}
and hence (ii) must be true.
Now from (i) and (ii), we have
$P_{\rm opt}^{\rm (sep)}(\gamma_0)-\gamma_1= (\Delta b + 2\Delta c)/4
\ge g(\hat{N})$.

For a positive product operator $\hat{G}=\hat{A}\otimes \hat{B}$
with $x(\hat{A})=x$ and $y(\hat{B})=y$,
we can bound $g$ by using the requirement for the positivity,
 $A_{11}\ge x^2 A_{00}$
and $B_{11}\ge y^2 B_{00}$, as 
\begin{eqnarray}
 g(\hat{G})\ge w(\hat{G})[(xy-\xi_0)^2
+(x-y)^2]/4.
\end{eqnarray}
Comparing it with Eq.~(\ref{eq:fpmG1}), we have  
$\eta_0w(\hat{G})g(\hat{G})\ge f_\pm(\hat{G})^2$ with 
$\eta_0\equiv 16[4+(1+\xi_0)^2]$. This relation
can be extended to any separable 
operator $\hat{K}=\sum_k \hat{G}_k$. In fact, with
$p_k\equiv w(\hat{G}_k)/w(\hat{K})$, we have 
$\eta_0w(\hat{K})g(\hat{K})
=\sum_k \eta_0 w(\hat{G}_k)g(\hat{G}_k)/p_k  
\ge \sum_k p_k [f_\pm(\hat{G}_k)/p_k]^2    
\ge [\sum_k f_\pm(\hat{G}_k)]^2
=f_\pm(\hat{K})^2$. 
Combining this with Eq.~(\ref{eq:fpmK2}), we have
\begin{eqnarray}
 g(\hat{K}_2^{(\pm)})&\ge& \frac{[w(\hat{K}^{(\pm)}_0)
+w(\hat{K}^{(\pm)}_2)]^2(1-\xi_0)^4}{\eta_0 w(\hat{K}_2^{(\pm)})}
\nonumber \\
&\ge& 4 \eta_0^{-1} (1-\xi_0)^4 w(\hat{K}^{(\pm)}_0)
\end{eqnarray}
Since $\hat{F}_2-(\hat{K}_2^{(+)}+ \hat{K}_2^{(-)})$ is 
positive, $\gamma_1\le P_{\rm opt}^{\rm (sep)}(\gamma_0)- 
g(\hat{K}_2^{(+)}+ \hat{K}_2^{(-)})$. Hence, using 
Eq.~(\ref{eq:kpmgamma}), we have a bound
$\gamma_1\le P_{\rm opt}^{\rm (sep)}(\gamma_0)
- 4 \eta_0^{-1} (1-\xi_0)^4 \gamma_0$ under LOCC,
namely,
\begin{eqnarray}
 && P_{\rm opt}^{\rm (LOCC)}(\gamma_0)\le u(\gamma_0)
\nonumber \\
&\equiv&
P_{\rm opt}^{\rm (sep)}(\gamma_0)-\frac{(1-2\gamma_0)^4\gamma_0}{4(1-\gamma_0)^2[1+4(1-\gamma_0)^2]},
\end{eqnarray}
which proves that there is a nonzero gap except $\gamma_0=0,1/2$.

We also give lower bounds on $P_{\rm opt}^{\rm (LOCC)}(\gamma_0)$
by finding specific LOCC protocols.
If Alice first measures on $\{\ket{+},\ket{-}\}$,
Bob's task is either the discrimination of 
$\{\ket{0},\ket{+}\}$ or that of $\{\ket{0},\ket{-}\}$,
and hence they can achieve $\gamma_1=l_1(\gamma_0)\equiv 
1- [2(1-\gamma_0)]^{-1}$. 
If Alice and Bob both measure on $\{\ket{0},\ket{1}\}$,
they achieve $(\gamma_0,\gamma_1)=(0,3/4)$.
Mixture of the 
$(\gamma_0,\gamma_1)=(0,3/4)$ protocol and the 
$(\gamma_0,\gamma_1)=(\sqrt{2}-1,l_1(\sqrt{2}-1))$
protocol gives another linear lower bound $l_2(\gamma_0)$. 
These are shown in Fig.~\ref{fig:1} (b).

To summarize, we have found a simple problem 
of state discrimination with 
$P^{(\rm glo)}_{\rm opt}>
P^{(\rm sep)}_{\rm opt}>P^{(\rm LOCC)}_{\rm opt}$. 
This means that, even for a qubit pair,
there exists a measurement that 
requires entanglement to execute, but never produces 
entanglement from scratch. The measurement is still 
useful, having a definite advantage over any LOCC measurement.
The pair of states in our example are 
very simple --- in fact, they are one of the simplest in a sense, 
considering that  
$P^{(\rm glo)}_{\rm opt}=P^{(\rm LOCC)}_{\rm opt}$ 
has already been proved 
for discrimination of 
any two pure product states \cite{Chen-Yang01}.  
This suggests that the existence of the gap may
not be rare but could show up in many other problems, 
all the more reason to investigate this striking 
irreversibility in more detail. The qubit pair 
is a convenient playground for such investigation, 
because operations are parametrized by fewer parameters 
and entanglement is understood more quantitatively. 
We hope that the present example would serve as a
driving force toward a better understanding about the
subtle boundary between quantum and classical interactions.

The authors thank K. Azuma for helpful 
discussions. 
This work was supported by a MEXT Grant-in-Aid 
for Young Scientists (B) 17740265.

\bibliographystyle{apsrev}
\bibliography{QI}

\end{document}